# Ordered community structure in networks

**Steve Gregory**
Department of Computer Science, University of Bristol, Bristol BS8 1UB, England

**Abstract.** Community structure in networks is often a consequence of homophily, or assortative mixing, based on some attribute of the vertices. For example, researchers may be grouped into communities corresponding to their research topic. This is possible if vertex attributes have discrete values, but many networks exhibit assortative mixing by some continuous-valued attribute, such as age or geographical location. In such cases, no discrete communities can be identified. We consider how the notion of community structure can be generalized to networks that are based on continuous-valued attributes: in general, a network may contain discrete communities which are ordered according to their attribute values. We propose a method of generating synthetic ordered networks and investigate the effect of ordered community structure on the spread of infectious diseases. We also show that community detection algorithms fail to recover community structure in ordered networks, and evaluate an alternative method using a layout algorithm to recover the ordering.

## 1. Introduction

Networks often exhibit a community structure, whereby groups of vertices (communities) exist such that the density of edges within a community is higher than average. In other words, an edge between any two vertices is more likely if the vertices are in the same community (or communities) than if they are not. Several types of community structure have been identified – disjoint, overlapping, hierarchical, etc. – but they all have the property that a network is partitioned into a finite number of (possibly overlapping) sets of vertices. This is true even in the case of fuzzy overlapping [1], where the membership of a vertex to a community may vary between 0 and 1. Numerous algorithms have been developed to detect communities; see Ref. [2] for an excellent review of many of them.

Perhaps the main reason for the formation of community structure in networks is homophily, or assortative mixing, by some attribute value(s) of the vertices: edges are more likely between vertices that have the same attribute value [3,4]. When we analyse a network, the vertex attributes are often unknown or missing, so all that remains is the induced community structure. For example, in a network of millions of Belgian mobile phone users, Blondel et al. [5] found three large communities, two of which correspond to the language (French or Flemish) spoken by the respective users. Studies of sexual networks have found many examples of assortative mixing: for example, partners are most likely to have the same race [6] and the same smoking status [7]. These characteristics correspond to a division of the population into a small number of large communities – e.g., the sets of all smokers and all non-smokers. Because each person is characterized by many attributes, he/she may belong to many overlapping communities, but we shall ignore the possibility of overlapping in this paper.

This correspondence between homophily and communities only holds if attribute values are discrete. However, many networks exhibit assortative mixing by continuous attribute values: the probability of an edge between two vertices is inversely related to the difference between their attribute values. An example of this which appears in most sexual networks is age. For example, Fig. 1 plots the distribution of age difference in 14833 sexual partnerships in which both partners are aged between 16 and 44, extracted from the NATSAL II dataset [8]. Some studies [9] have even found that the geographical distance between partners' residences within a city is shorter than expected by chance. This continuous kind of mixing does not give rise to community structure in the usual sense, but it is fundamentally similar, so we call it *continuous community structure*.

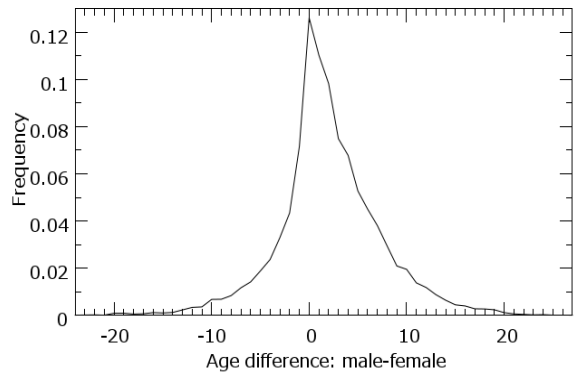

Fig. 1. Age difference (male-female) between sexual partners, from NATSAL II dataset. Mean difference is 2.02 years.

Continuous attributes can be "stratified" into a finite number of ranges. This is very commonly done with attributes such as level of education, social class/income, and alcohol consumption, all of which have been found to be assortatively mixed in sexual network surveys [6,7,8]. Here, there is a relatively high probability of an edge between two vertices whose attribute values fall into the same range. This gives rise to community structure, but with an additional property: intercommunity edges are more likely between vertices in adjacent communities (communities corresponding to ranges adjacent to each other) than between vertices in more distant communities. We call this *ordered community structure*.

Continuous and ordered community structure are closely related. Even when attributes are continuous in the real world, they are often stratified in collected data; for example, the continuous data plotted in Fig. 1 is actually recorded in integer form. We can think of continuous community structure as the limiting case of ordered community structure in which communities are very small. Therefore, in subsequent sections we shall simply refer to both types as *ordered*. Following Newman [3,4], we shall refer to both continuous and ordered discrete attributes as *scalar* attributes.

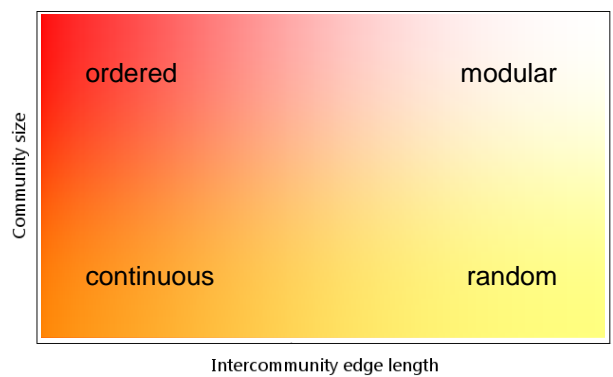

Fig. 2. Continuous and ordered community structure.

Fig. 2 shows our classification of types of community structure. The normal, "modular", form of community structure corresponds to a large "intercommunity edge length"; that is, an intercommunity edge has an equal probability of linking to *any* other community. As communities become small, the network becomes "random", for example, like an Erdős-Rényi network. We can obtain "ordered" and

"continuous" community structure by reducing the average intercommunity edge length, so that most edges between communities will tend to connect nearby communities (those with similar attribute values).

Although community structure has been a subject of intense research in recent years [2], ordered structure has been almost completely ignored. A notable exception is the work by Expert et al. [10], who propose a way to factor out assortative mixing by continuous attributes so that the (unordered) discrete communities can be identified more easily, in networks that have both types of attribute. In contrast, we focus on analysing networks with only ordered attributes, and recovering the ordering rather than discarding it.

In the remainder of this paper we consider several aspects of networks with ordered community structure. First, in Section 2, we propose a method of constructing synthetic networks with these forms of community structure. Section 3 describes the effect of ordering on the transmission of a disease through a network. In Section 4 we show that conventional community detection algorithms fail to work well on ordered networks, while Section 5 discusses how to overcome this problem and even recover the correct order of communities.

## 2. Constructing artificial networks with ordered community structure

Synthetic networks are very important in the study of network processes and algorithms. For example, community detection algorithms can be evaluated more effectively on synthetic networks than on real networks because we can easily vary the parameters and compare the results with the expected results. Synthetic social networks are also ubiquitous in studies of disease transmission, because it is difficult or impossible to obtain a complete real network that reflects properties of the disease being studied. For example, a study of TB requires a network of respiratory contacts of the appropriate form for transmitting the disease, and this information is not readily available.

The *benchmark* network generator of Lancichinetti et al. [11], which we shall call the LFR method, has become a very widely used tool for constructing networks with community structure. It produces networks that are claimed to possess properties found in real networks, such as heterogeneous distributions of degree and community size, and can be controlled by several parameters. Some of the parameters specify properties of communities: $N$ (number of vertices), $c_{min}$ and $c_{max}$ (minimum and maximum community size), and $\tau_2$ (exponent of the power-law distribution of community sizes). The other parameters specify properties of the generated network: $\langle k \rangle$ (average degree), $k_{max}$ (maximum degree), $\mu$ (mixing parameter: each vertex shares a fraction $\mu$ of its edges with vertices in other communities), and $\tau_1$ (exponent of the power-law distribution of vertex degrees).

One method of generating networks with continuous community structure is that of Read and Keeling [12], which we shall call the RK method. This has just three parameters: $N$ (number of vertices), $\langle k \rangle$ (average degree), and $D$ (average distance between vertices). Vertices are given positions in 2-dimensional space, uniformly distributed across a plane of size $\sqrt{N} \times \sqrt{N}$. The probability of an edge between two vertices is $p.\exp(-d^2/2D^2)$, where $d$ is the Euclidean distance between the two vertices and $p$ is chosen to make the average degree equal to $\langle k \rangle$. When $D$ is large (at least $\sqrt{N}/2$), the method generates an Erdős-Rényi network. As $D$ is reduced to 1 (or slightly less) the network structure becomes increasingly "local": vertices connect mainly to nearby ones and the clustering coefficient increases.

The RK method can be generalized to a different number of dimensions, although this was not suggested in Ref. [12]. For example, a 1-dimensional variant would be more suitable for representing assortative mixing by age in a social network.

Using the terminology of Fig. 2, the RK method can generate "random" and "continuous" networks, but it does not feature community structure and it generates networks with a poisson degree distribution. Conversely, the LFR method generates networks with a power law degree distribution that exhibit "random" or "modular" structure, depending on the community size parameter, but cannot produce "ordered" or "continuous" networks. This is because intercommunity edges are placed arbitrarily between different communities.

We have implemented a network generator that can produce all four types of network structure. It is based on the LFR method but can also impose an order on the communities. It comprises three phases:

1. Use the LFR method to construct a set of communities and network satisfying parameters $N$, $c_{min}$, $c_{max}$, $\tau_2$, $\langle k \rangle$, $k_{max}$, $\mu$, and $\tau_1$.
2. Assign each community a position on a $d$-dimensional grid of size $g$, where $g$ is the number of communities. E.g., for two dimensions, a $\sqrt{g} \times \sqrt{g}$ grid is used.
3. Repeatedly "rewire" edges to reduce the average length of intercommunity edges to $D$.

In step 2, instead of randomly placing communities, we actually use integer coordinates $(x,y)$ such that $0 \leq x,y < \lceil \sqrt[d]{g} \rceil$, because this ensures that the minimum possible value of $D$ is 1.

In step 3, we choose any two edges $\{u,v\}$ and $\{x,y\}$ such that $u$, $v$, $x$, and $y$ belong to four distinct communities, $c_u$, $c_v$, $c_x$, and $c_y$. If $d(c_u,c_y) + d(c_v,c_x) < d(c_u,c_v) + d(c_x,c_y)$, where $d(c_1,c_2)$ denotes the Euclidean distance between $c_1$ and $c_2$, we replace these two edges by $\{u,y\}$ and $\{v,x\}$. Alternatively, if $d(c_u,c_x) + d(c_v,c_y) < d(c_u,c_v) + d(c_x,c_y)$, we replace the two edges by $\{u,x\}$ and $\{v,y\}$. This step reduces the average intercommunity edge length without changing the degree of any vertices.

Fig. 3 illustrates the edge rewiring procedure. Fig. 3 (top left) shows a network with 6 communities generated by the LFR method, placed in 2-dimensional space. We arbitrarily choose intercommunity edges $\{22,32\}$ and $\{34,35\}$, which have length 2.24 ($= \sqrt{(2^2+1^2)}$) and 1 ($= \sqrt{(0^2+1^2)}$), respectively. In Fig. 3 (top right) these are replaced by $\{22,35\}$ and $\{24,34\}$ which both have length 1. This reduces the average intercommunity edge length from 1.47 to 1.41. After further iterations, the average is reduced to 1.04, as shown in Fig. 3 (bottom), in which only two edges are longer than 1.

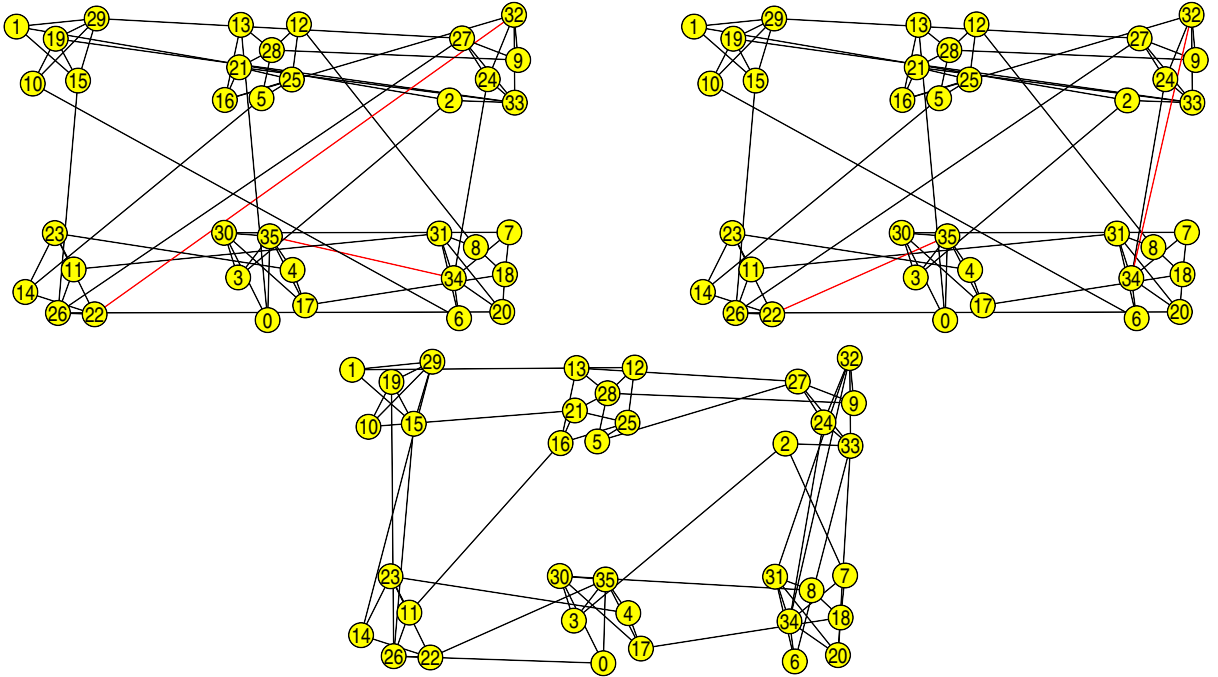

Fig. 3. Constructing ordered networks by rewiring edges. Top left: unordered LFR network. Top right: network after swapping two edges of network, shown in red. Bottom: network after several more rewiring steps.

Our method can easily be generalized to different numbers of dimensions. In subsequent sections we use our network generator to produce 2-dimensional and 1-dimensional networks. We experiment with all four types of network shown in Fig. 1, varying the community size from the range 6-12 ("small" communities) to 140-280 ("large" communities) and the average intercommunity edge length from 1.1 ("ordered") to $D_R$ ("unordered"), where $D_R$ is the average intercommunity edge length of the original unordered network. We normalize the average intercommunity edge length by dividing it by $D_R$, so that our results are independent of the number of communities, which in turn depends on the community size. All experiments use the parameters $N$=10000, $\tau_2$=1, $\langle k \rangle$=10, $k_{max}$=25, $\tau_1$=2, and $c_{max}$ is always $2 \times c_{min}$. Results plotted are the average of 100 runs.

Fig. 4 shows that the clustering coefficient (the fraction of triples that are "closed") can be increased by either increasing orderedness (reducing $D/D_R$) or reducing community size ($c_{min}$ and $c_{max}$). In both cases, the reason for the increased clustering is the smaller number of vertices to which each vertex is likely to be linked by an edge.

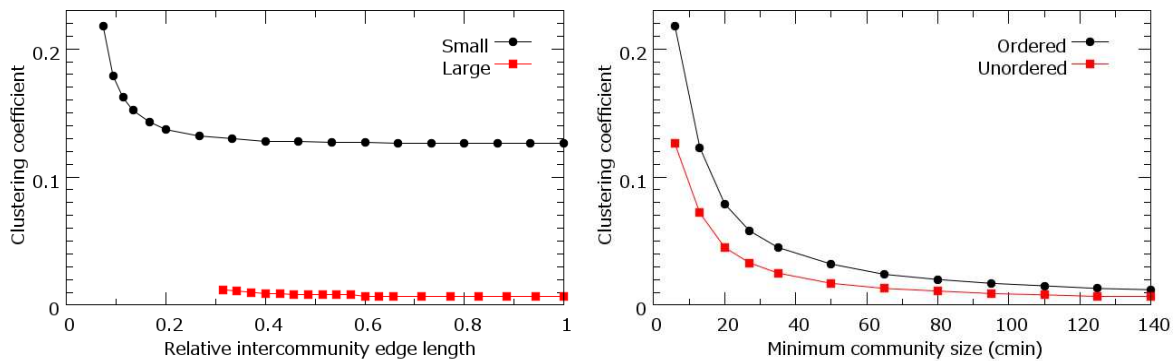

Fig. 4. Effect of network structure on clustering coefficient. Left: as intercommunity edge length varies. Right: as community size varies. These results are for 2-dimensional networks with $\mu=0.5$.

### 3. Disease transmission in networks with ordered community structure

Disease modellers study the spread of a disease through a network that represents the contacts relevant to the particular disease. It is well known that the spread of a disease can be strongly influenced by network structure. In particular, many disease models have assumed an ordered network structure; for example Mills et al. [13], which uses the RK model [12] to simulate local respiratory contacts, and Turner et al. [14], which uses assortative mixing by age to form sexual contacts. To a lesser extent, the effect of discrete, unordered community structure has also been studied [15]. In this section, we examine the effect of our four types of network structure on disease transmission.

We simulate the spread of an infection using a simple susceptible-infected-resistant (SIR) model in which, at each time step, there is a fixed probability (0.08) of the disease being transmitted to a susceptible vertex from each infectious neighbour, and a vertex remains infectious for a fixed period (four time steps). We vary community size and average intercommunity edge length and measure the epidemic duration (number of time steps from beginning to end), peak prevalence (maximum fraction of infectious vertices at any time), and final size (total number of vertices infected).

Results for 2-dimensional networks with $\mu=0.3$ are shown in Fig. 5. As the intercommunity edge length is reduced (Fig. 5 (left)), the duration increases and peak prevalence decreases. The same happens as community size is reduced (Fig. 5 (right)). In other words, the infection travels more slowly in networks with ordered communities, and especially small ordered communities. In contrast, when communities are unordered, the network has the "small world" property and the infection can spread rapidly. The final epidemic size remains largely unaffected by the type of communities.

### 4. Community detection in networks with ordered community structure

As explained in Section 1, "modular" and "ordered" networks possess a discrete community structure but "continuous" and "random" networks do not. In this section we investigate whether community structure can be found in our different forms of network. To do this, we use the Infomap algorithm [16], which is sometimes regarded as one of the best current algorithms for detecting disjoint communities [17]. We generate 2-dimensional networks with $\mu=0.5$ and varying community size and intercommunity edge length, detect communities using Infomap, and compare the communities found with the actual communities using the normalized mutual information (NMI) measure [18].

Fig. 6 (left) shows the results of varying intercommunity edge length. For unordered networks, Infomap works well, but as the network becomes more ordered, the performance suddenly drops. This is true for both small and large communities. The reason is that, as more intercommunity edges fall between neighbouring communities, they appear to merge and cannot be distinguished. Fig. 6 (right) confirms that the community size has little effect; except that Infomap works slightly less well for larger communities.

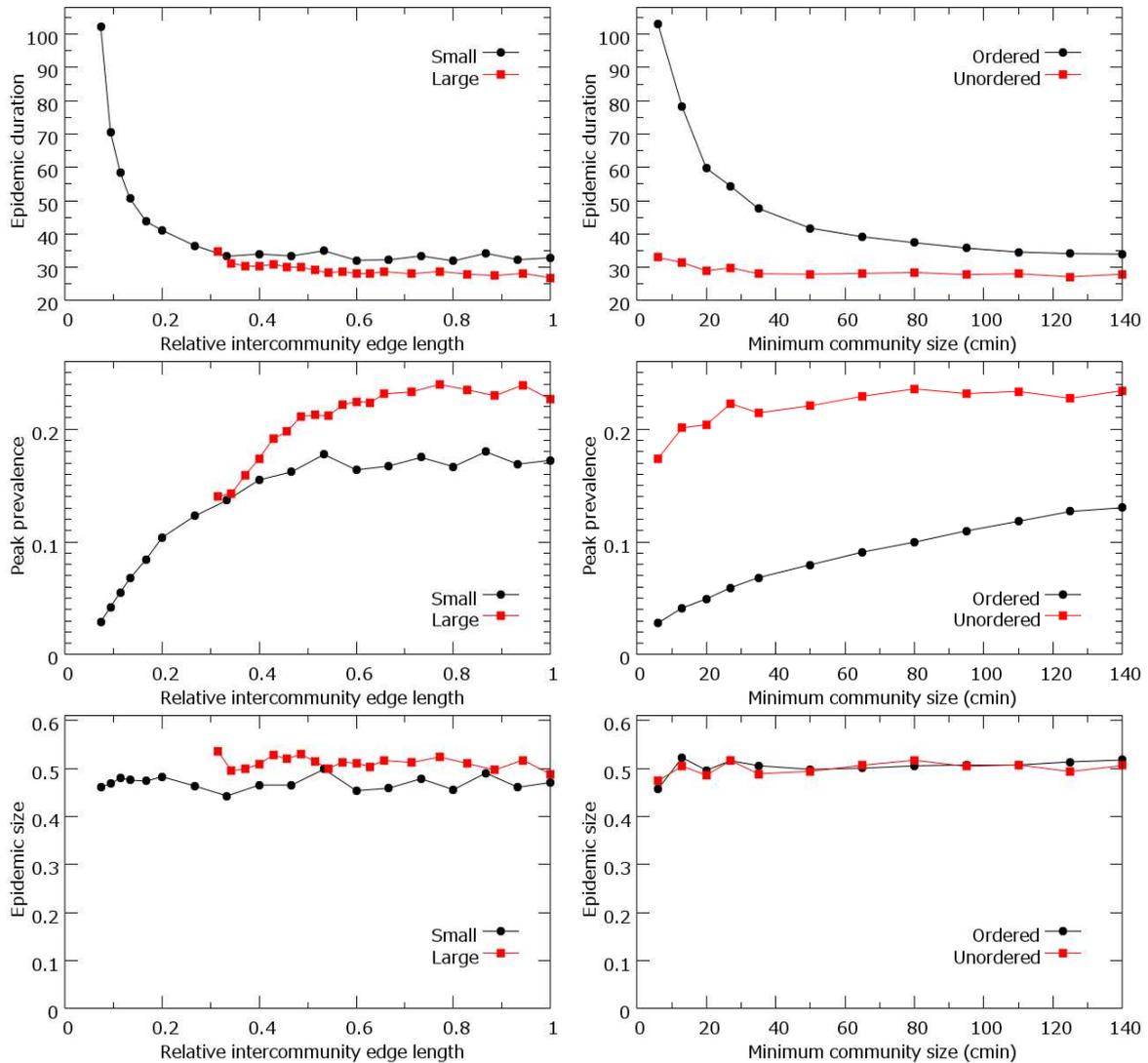

Fig. 5. Effect of network structure on disease dynamics. Left: as intercommunity edge length varies. Right: as community size varies. Top: epidemic duration. Centre: peak prevalence. Bottom: final size of epidemic. These results are for 2-dimensional networks with $\mu=0.3$.

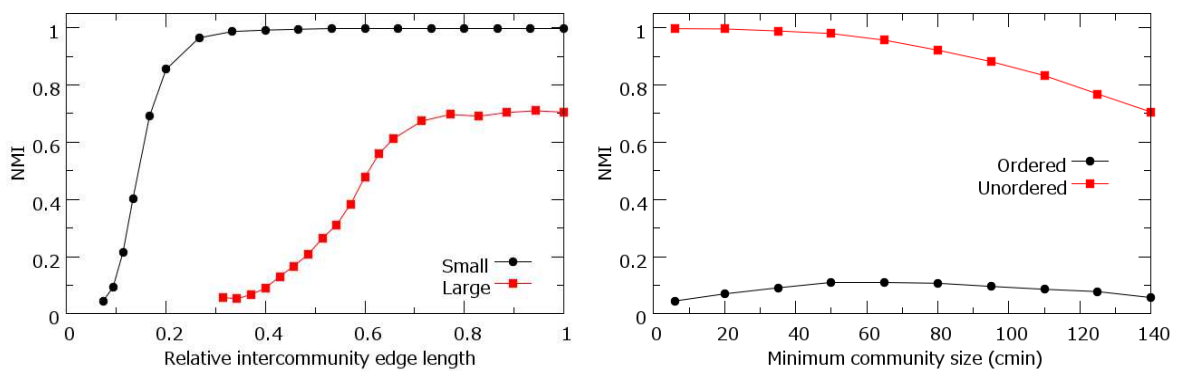

Fig. 6. Effect of network structure on community detection. Left: as intercommunity edge length varies. Right: as community size varies. These results are for 2-dimensional networks with $\mu=0.5$.

## 5. Recovering ordered communities

Our results in Section 4 suggest that community detection algorithms are unable to find communities if they are ordered; this is confirmed by experiments with other algorithms. This raises two questions:

1. Are there alternative methods that can detect communities in an ordered network?
2. If so, is it possible to recover the ordering of communities, or the ordering of vertices in a "continuous" network?

### 5.1. Layout algorithms

Recovering the "order" of vertices or communities from a network is equivalent to assigning each vertex or community to a position in some metric space. This task is already performed by *layout algorithms*, which are usually used for visualization purposes. Noack [19] has elegantly demonstrated that force-directed layout algorithms can be used to detect communities, by placing their vertices in similar locations, but he did not consider the relative positions of vertices in different communities because he was not investigating ordered networks. Our hypothesis is that layout algorithms may be useful for recovering ordering information from networks.

We have experimented with Noack's LinLogLayout algorithm [20] on ordered networks. The algorithm calculates positions of vertices by modelling attractive and repulsive forces between vertices. Vertices joined by an edge $\{u,v\}$ are attracted by a force $d(u,v)^a$, while all vertex pairs $\{u,v\}$ are subject to a repulsive force $d(u,v)^r$, where $d(u,v)$ is the distance between (the current positions of) $u$ and $v$. We use the default parameters, $a=1$, $r=0$.

Fig. 7 shows the 2-dimensional layout computed for a small network ($N=1000$) with $\mu=0.3$, $c_{min}=20$, $c_{max}=40$, and an average intercommunity edge length of 1.1. Each vertex is represented by a coloured dot, the colour representing a community identifier decided by the algorithm. The network used has 36 communities, which were assigned positions on a 6×6 grid (coordinates (0,0) to (5,5)) before the network was constructed. Noack's program makes intracommunity edges very short, so that each community's vertices are located together. In the figure, we have manually labelled communities with their original coordinates. This shows that the layout algorithm is remarkably accurate in its relative positioning of the communities.

### 5.2. Assessing layout quality

In order to quantitatively measure the quality of a computed layout $l_l$ of a network $G$ ($=(V,E)$), we need to compare it with the "correct" layout $l_o$ (the one from which the network was constructed). To compare $l_l$ and $l_o$, we first scale both so that all coordinates cover the range [0,1], and then measure the difference between them, $D(l_l,l_o,G)$. There are various ways to define $D$, the simplest being the average distance between the absolute position of each vertex in $l_l$ and its position in $l_o$:

$$D_{absolute}(l_1,l_2,(V,E)) = \frac{1}{n}\sum_{v \in V} d(v_{l_1}, v_{l_2}), \quad (1)$$

where $v_l$ denotes the position of vertex $v$ in layout $l$ and $d(p,q)$ denotes the Euclidean distance between positions $p$ and $q$.

However, in general, a layout algorithm can only recover relative positions, not absolute positions, of vertices. Therefore, it may be more realistic to compare the distance between pairs of vertices in each layout. The following measure averages this over all pairs of vertices:

$$D_{global}(l_1,l_2,(V,E)) = \frac{1}{n(n-1)}\sum_{u \in V}\sum_{v \in V} \left| d(u_{l_1}, v_{l_1}) - d(u_{l_2}, v_{l_2}) \right|. \quad (2)$$

Alternatively, a local version of the same measure can be computed, averaging over only the pairs of vertices that are linked by an edge:

$$D_{local}(l_1,l_2,(V,E)) = \frac{1}{|E|}\sum_{\{u,v\} \in E} \left| d(u_{l_1}, v_{l_1}) - d(u_{l_2}, v_{l_2}) \right|. \quad (3)$$

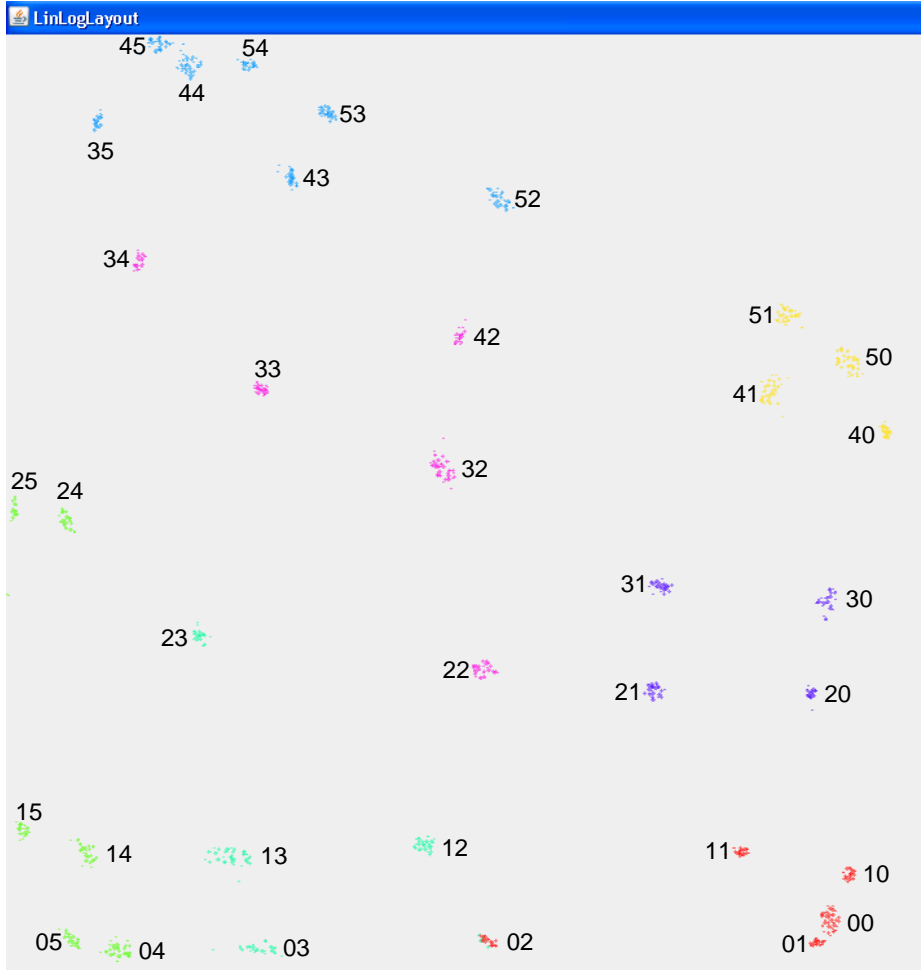

Fig. 7. Layout computed by LinLogLayout algorithm for an ordered 2-dimensional network with 36 communities.

$D(l_l,l_o,G)$ varies between 0 and $\sqrt{d}$ for a $d$-dimensional layout. However, even for a random layout $l_r$, in which vertices are uniformly distributed, $D(l_r,l_o,G)$ will be much less than the maximum possible, so we divide the actual difference by that of a random layout. Our measure of "position quality", $P$, is therefore defined as:

$$P = 1 - \frac{D(l_l,l_o,G)}{D(l_r,l_o,G)}, \qquad (4)$$

which ranges from 0 (if $l_l$ is a random layout) to 1 (if $l_l$ is perfect). This gives three versions of position quality: $P_{absolute}$, $P_{global}$, and $P_{local}$.

*5.3. Experiments on synthetic networks*

Fig. 8 shows results of experiments on 2-dimensional synthetic networks with $N=10000$ and $\mu=0.3$. We vary the community size or the intercommunity edge length, as before, and plot the local and global position quality of the computed layout $l_l$. We also plot the local position quality of a special layout obtained from a partition of the network by assigning a random position to each community and using that position for all vertices in that community. Although this layout contains no information about intercommunity edge length, it yields the correct length (0) for all intracommunity edges. The partition used is the one from which the network was constructed, so these results represent an upper bound on the performance obtainable if the partition were found by a community detection algorithm.

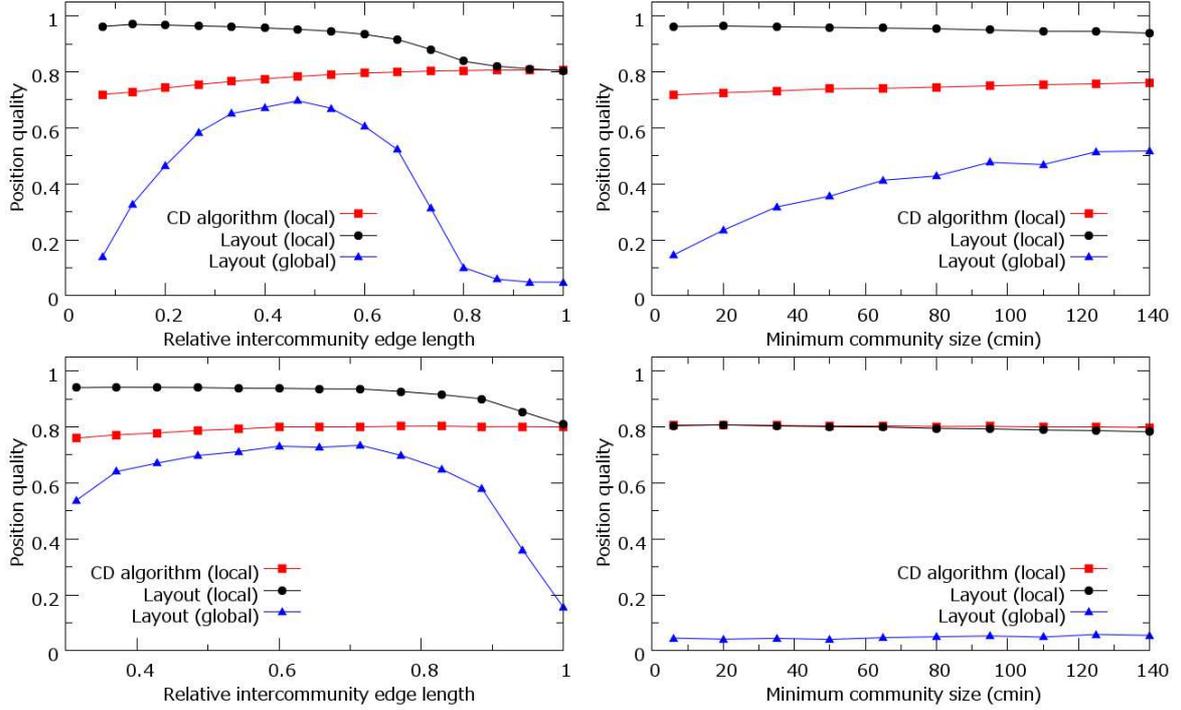

Fig. 8. Recovering vertex positions from a 2-dimensional network ($\mu$=0.3). Top left: small communities. Bottom left: large communities. Top right: ordered networks. Bottom right: unordered networks.

For local position quality, the layout algorithm performs much better than the community detection algorithm when intercommunity edge length is small (ordered networks). Only when the network becomes completely unordered do the layout results and community detection results converge (Fig. 8 (left)). As community sizes increase (Fig. 8 (right)), the community detection results become very slightly better than the layout algorithm's results. This is because, when communities are large, the layout algorithm is forced to use longer intracommunity edges, while the community detection algorithm always uses edges of length 0.

The global position quality is best in the middle of the range of intercommunity edge length. It is inevitably low when intercommunity edge length is high (unordered networks), because only the relative positions of vertices in the same community can be preserved, and is also low when the network is highly ordered, especially when communities are small.

Fig. 9 shows results for the same networks laid out in one dimension. Their characteristics are the same as the results of Fig. 8. This time we have also plotted the absolute position quality of the computed layout after a simple modification: we calculate the average position in $l_l$ of all vertices with position in $l_o$ that are < 0.5 and ≥ 0.5, respectively, and reverse the order of $l_l$ if the first is greater than the second. The absolute position quality measure varies in the same way as the global one. Both of these remain above 0 even for unordered networks, because the layout algorithm keeps members of each community together.

*5.4. Experiments on real networks*

To evaluate how well the technique can work in practice, we have applied it to three real networks with ordered structure.

Our first example is the college football network [21], which is one of the most widely studied examples of a community-structured network. It contains 115 vertices representing football teams and 613 edges representing games between them. The teams are grouped into disjoint "conferences", most of which are based in a particular region of the United States. This constitutes a natural community structure because most games are played between teams in the same conference. As Girvan and Newman [21] noted, inter-conference games are not uniformly distributed but are more common between teams that are geographically close to each other. In other words, the network has a 2-dimensional ordered community structure. Although this network is often used as a benchmark for community detection algorithms, the order (positions) of the conferences has been ignored until now. As shown in Fig. 10, a layout algorithm can recover their relative positions remarkably well.

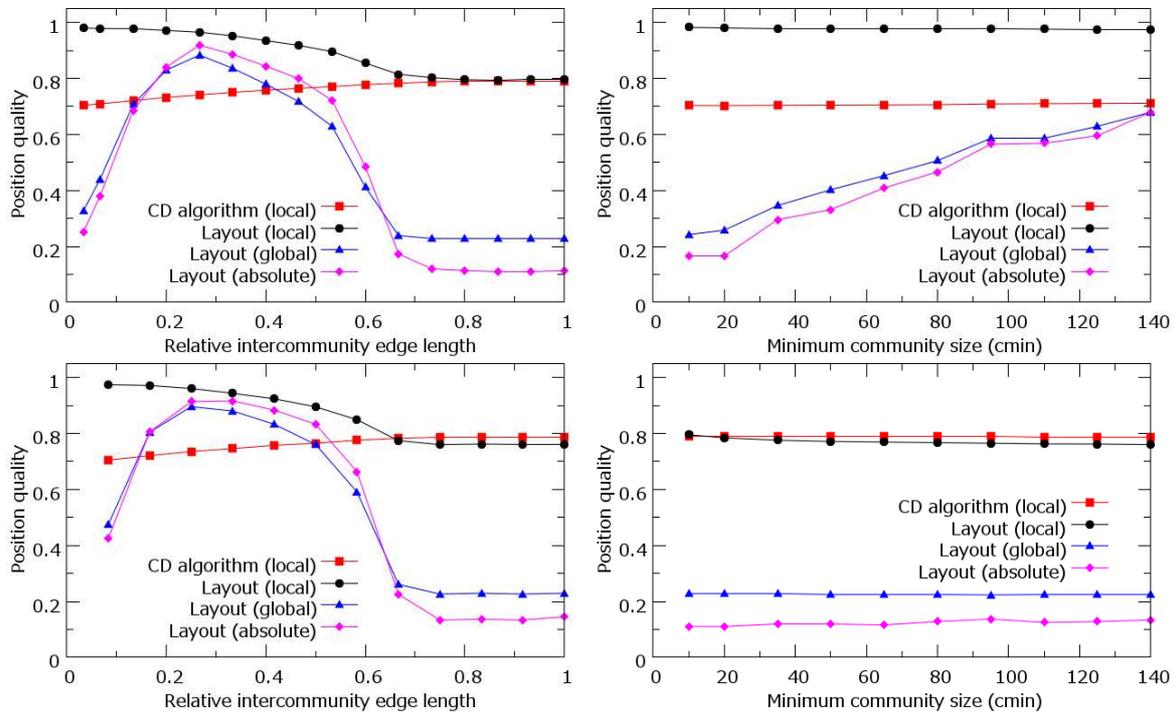

Fig. 9. Recovering vertex positions from a 1-dimensional network ($\mu$=0.3). Top left: small communities. Bottom left: large communities. Top right: ordered networks. Bottom right: unordered networks.

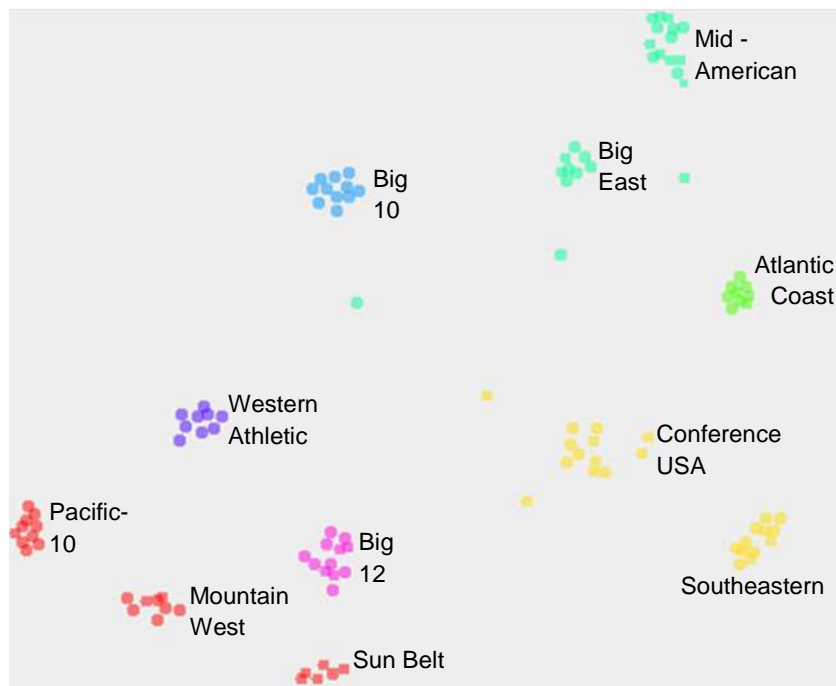

Fig. 10. Layout computed by LinLogLayout for college football network. Each dot is a football team and the groups of dots are communities, labelled with the name of the corresponding "conference".

Our next example is also a 2-dimensional spatial network, but without community structure. This is a network of residence-workplace travel flows taken from the 2001 UK census [22]. Vertices represent local authority areas in England, Wales, and Scotland, while weighted, directed edges represent commuting trips. The original network contains 381 vertices and 26023 edges. We removed self-edges and replaced each pair of directed edges $(u,v)$ and $(v,u)$, with weights $w_1$ and $w_2$, by an undirected edge $\{u,v\}$ with weight $w_1+w_2$. Finally, we deleted edges with weight less than 700, resulting in a network with 372 vertices and 1928 edges. A threshold above 720 would cause a split into two components, while a small threshold would retain all 381 vertices but too many insignificant edges. We used the MapIt web service [23] to find the centroid of each local authority area, as a cartesian grid reference, and used these positions as the ground truth.

Our hypothesis was that the travel network would be assortatively mixed by geographical location; in other words, commuting trips tend to be short. This is confirmed by Fig. 11, which plots the frequency distribution of trip length; more exactly, the distances between locations that are the endpoints of at least 700 trips. The reason for the low frequency of very short trips is that these are mostly trips within a single area, which we have excluded by removing self-edges.

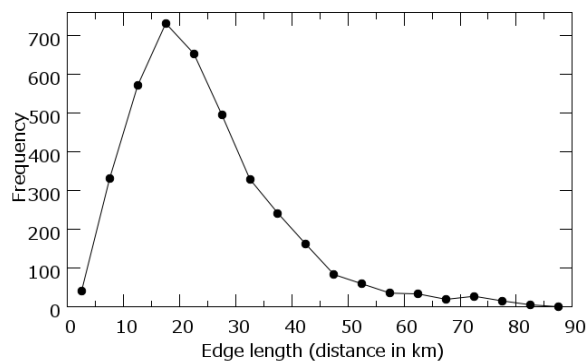

Fig. 11. Travel network. Frequency distribution of distances between locations with heavy travel flow. The mean is 24.7km and the maximum possible distance between locations is 824km.

In our layout experiment, the local and global position quality was 0.914 and 0.310, respectively. The reason for the low global quality is that most traffic flow is to or from a few large cities, especially London, while there are fewer edges between the less important areas. This means that insignificant areas are positioned well relative to large cities, but not so well relative to each other.

Our final example is a network of tennis games taken from a website [24] that arranges games by seeking to match players with similar ability, as determined by the players' ratings. Ratings are on the standard National Tennis Rating Program scale (1-7), and are updated dynamically by the system based on the results of the games that it has arranged. Our network has a vertex for each player and an edge between each pair of players that have played at least one (singles) game during the time period concerned, which is the year ending June 11, 2011. This network has 443 vertices and 1404 edges. The attribute of each vertex is the player's rating at the end of this time period. Because of the way in which games are arranged, we expect that the network will be assortatively mixed by player rating; i.e., it will be a 1-dimensional ordered network. However, this is complicated by the dynamic nature of the ratings: a player's final rating may differ from that at the time when a game was arranged. Nevertheless, the difference between matched players' ratings is indeed quite small, as shown in Fig. 12.

The layout algorithm produces a solution whose local and global position quality are 0.872 and 0.611, while its absolute position quality is 0.311. However, it is possible to improve the quality if the real ratings of a few ($k$) vertices are known. We ran the layout algorithm 20 times, measured the average difference between the real ratings and computed ratings for these $k$ vertices each time, and chose the layout with the smallest average difference. The $k$ vertices were randomly chosen on each run, and results were averaged over 100 runs. This method improved the absolute position quality, as shown in Fig. 13, increasing to 0.574 for $k$=20. What this means in practice is that we are able to predict the rating of *all* players with an average error of 0.333, compared with 0.781 for a random guess. This is almost as good as the best predictions possible even with complete information about the ratings. For example, if we predict a player's rating to be the same as a randomly chosen neighbour whose rating is known, the average error would be 0.196; if we assume that all neighbours' ratings are known, and use the mean of these ratings, the error would be 0.149.

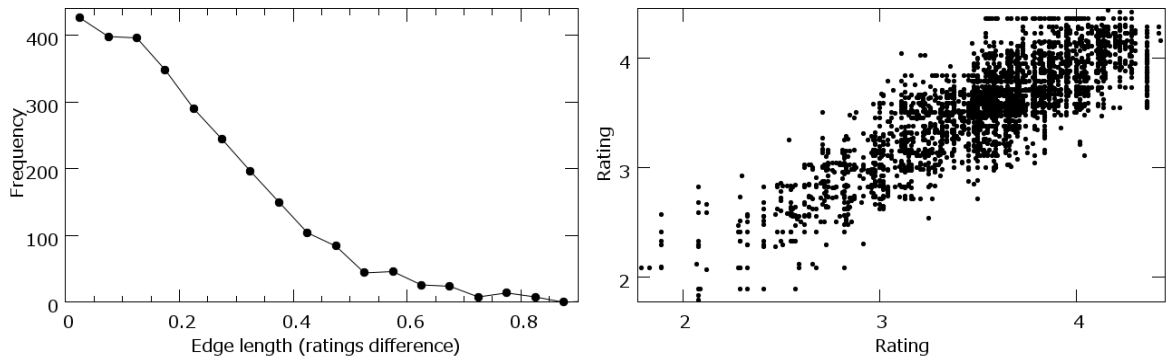

Fig. 12. Tennis network. Left: frequency distribution of differences in rating between players in each game. The mean is 0.21 and the maximum possible difference for this set of players is 2.64. Right: scatter plot of games, where the *x* and *y* coordinates show the rating of each player.

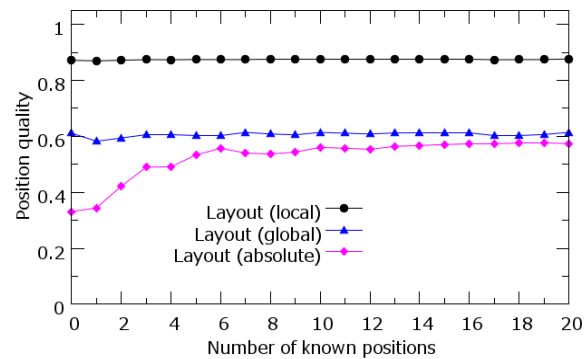

Fig. 13. Tennis network. Results of choosing the best of 20 runs that minimizes the average error between computed position and known position of a certain (varying) number of vertices.

Incidentally, it is interesting to compare this result with the work of Radicchi [25], who also used network analysis to rank tennis players, but using the results of games. Our method does not use results at all, but relies only on the tendency of opposing players to have similar ability. On the other hand, our method would not work for the network of Ref. [25] because it covers a 42-year period, so players are assortatively mixed by time as well as ability.

## 6. Conclusions

Community structure admits a convenient mesoscopic view of a network which can be easier to understand than the microscopic view as a set of vertices. Perhaps for this reason, community structure in networks has been intensively studied in recent years. It has gradually become clear that networks possess many different types of community structure: for example, overlapping and hierarchical communities as well as disjoint ones. However, the possibility that communities may be ordered has been largely neglected until now, despite the fact that scalar vertex attributes are very common in real networks. Indeed, such attributes are ubiquitous in the contact networks used for modelling infectious diseases, which inspired this work.

Ordered community structure shares some properties with conventional discrete community structure. For example, both cause an increase in a network's clustering coefficient (transitivity), as we showed in Section 2. However, there are also differences, which we have examined in this paper. First, an infectious disease travels more slowly in a network with ordered communities, especially small ones, than with unordered communities, which has the small-world property (Section 3). Second, conventional community detection methods cannot be used when communities are ordered (Section 4), and so new algorithms are needed. As a starting point, in Section 5, we have shown that one existing layout algorithm works well for this purpose. Our final contribution (Section 2) was to present an algorithm for constructing networks with ordered community structure, by refining the widely-used LFR method.

One area for future work is to develop alternative algorithms for recovering ordering from ordered networks. Other layout algorithms might be suitable for this purpose. Another topic is to analyse networks that have assortative mixing by more than one continuous attributes (e.g., age and weight), to find the correct order of both. A network with two 1-dimensional attributes is different from one with one 2-dimensional attribute, in general. This problem is analogous to the problem of detecting overlapping communities, but in the context of continuous, instead of discrete, attributes.